\newcommand{\CLASSINPUTtoptextmargin}{0.75in}
\newcommand{\CLASSINPUTbottomtextmargin}{1in}
\documentclass[conference,10pt]{IEEEtran}
\IEEEoverridecommandlockouts

\usepackage{amsmath,amssymb,amsfonts}
\usepackage{algorithmic}
\usepackage{graphicx}
\graphicspath{{./Figures/}}
\usepackage{textcomp}
\usepackage{xcolor}
\usepackage{subfiles} 
\usepackage{subfig}
\usepackage{color,soul}
\usepackage{cite}
\usepackage{amsmath}
\usepackage{multirow}
\usepackage{makecell}
\usepackage{hyperref}
\usepackage{fancyhdr}

\def\BibTeX{{\rm B\kern-.05em{\sc i\kern-.025em b}\kern-.08em
    T\kern-.1667em\lower.7ex\hbox{E}\kern-.125emX}}
\fancypagestyle{firststyle}{
	\fancyhf{}
	\fancyhead[L]{K. F. Nieman, O. Kanhere, and S. S. Ghassemzadeh, ``cmWave/FR3 Large-Scale Channel Characterization for Urban Macro/Micro and Suburban Environments,'' \textit{in GLOBECOM 2025 - 2025 IEEE Global Communications Conference, Taipei, Taiwan, Dec 2025, pp.
1–6. } }    %

}    
\begin{document}
\bstctlcite{IEEEexample:BSTcontrol}

\title{cmWave/FR3 Large-Scale Channel Characterization for Urban Macro/Micro and Suburban Environments
}
\author{\IEEEauthorblockN{Karl F. Nieman, Ojas Kanhere, and
Saeed S. Ghassemzadeh}
\IEEEauthorblockA{AT\&T Labs, Austin, TX USA\\
Email: \{kn2644, ok067s, sg2121\}@att.com
}}
\maketitle
\thispagestyle{firststyle}

\begin{abstract}
This study delves into the comprehensive characterization of large-scale channels at centimeter wave frequencies 7-15 GHz for urban macro/micro and suburban environments. Path-loss, large-scale fading, and angular channel statistics are presented. Urban environments exhibited higher path loss and delay spread due to dense obstacles, whereas suburban areas showed relatively lower path loss but significant variability due to fewer but larger obstructions. The findings provide valuable insights for network planners and engineers, aiding in the development of more efficient and adaptive communication strategies. Enhanced models for channel prediction and system design are proposed, contributing to the advancement of next-generation wireless networks.

\end{abstract}


\section{Introduction}
\label{sec:intro}
The rapid advancements of wireless communication technologies, particularly with the advent of 5G and the impending development of 6G networks, necessitates a thorough understanding of channel characteristics across various frequency bands. Centimeter-wave (cmWave) frequencies, specifically within Frequency Range 3 (FR3: 7-24 GHz), offer substantial bandwidths that can support high data rates and low latency required for advanced applications. Like other frequency bands, the propagation behavior at these frequencies is influenced by environmental factors such as building density, foliage, and terrain variations, especially in diverse urban macro/micro and suburban settings. This complexity necessitates a multi-faceted approach to channel characterization to accurately capture the nuances of signal behavior in these environments. 

The 3GPP Spatial Channel Model (SCM) \cite{3gpp.38.901} is a widely utilized model for radio access networks system design, capturing essential channel characteristics such as large-scale and small-scale fading, delay spread, and other critical parameters. Originally developed for Frequency Range 1 (FR1:frequencies up to 6 GHz), the SCM was subsequently extended to include Frequency Range 2 
 frequency bands(FR2:frequencies 24.25-52.6 GHz) and even frequencies up to 100 GHz.  However, due to lack of empirical data at FR3 bands and variability of propagation characteristics at actual deployment scenarios the extrapolation of parameters from lower or higher frequency bands for FR3 may be inadequate. Therefore, it was found  essential to develop new channel parameters based on measurement campaigns conducted in real deployment scenarios in FR3 frequency bands. Insights gained from such comprehensive characterization are critical for optimizing network planning, improving signal coverage, enhancing data throughput, and ensuring the reliability of future wireless communication systems. These efforts will ultimately support the successful deployment and operation of 5G and 6G networks, enabling them to meet the high-performance demands of emerging technologies and applications.

Consequently, numerous research institutions \cite{Kim_2015, Nguyen_2016_UMa,Roivainen_2017, Kristem_2018, Shakya_2024_outdoor} have initiated efforts to characterize wireless channels in the FR3 bands to address these challenges and enhance the model’s applicability. However, there is still a significant need for more detailed and comprehensive studies in realistic deployment scenarios.  It should be noted that the gap in FR3 outdoor channel models primarily revolves around the limited understanding and insufficient empirical data on how signals propagate at these higher frequencies, particularly in diverse real-world environments. In this paper, we attempt to address this gap by conducting an extensive measurement campaign across various deployment scenarios and reporting on the channel parameters computed from these measurements. We investigate three scenarios of interest: urban micro cell (UMi), urban macrocell (UMa), and suburban macrocell (SMa).

The remainder of this paper is organized as follows: In Section \ref{sec:HW_Experiments} we describe the experimental equipment, scenarios, and data processing methods. Section \ref{sec:KeyFindings} covers our key findings of the channel parameters followed by conclusions and future work in Section \ref{sec:conclusions}.

\section{Measurement Hardware, Experiment Scenarios and Database}
\label{sec:HW_Experiments}

For our outdoor measurement campaign, a tripod-mounted transmitter is positioned atop rooftops or other structures, such as parking garages, at heights between 10 and 35 meters above the street level.  This antenna height ensures compliance with the deployment scenario specification. The transmitted signal is simultaneously received via two different antenna configurations: 1) two omnidirectional antennas and 2) four phased array antennas as described in \cite{Nieman_2025_01}. The receiving antennas are mounted atop a van that is driven along each street throughout the planned coverage area. In total, four sites were selected: one for the UMi scenario, one for the UMa scenario, and two for the SMa scenarios. At each site, the transmit antenna was directed toward two to three distinct pointing directions, creating multiple non-overlapping sectors. Data collected from these sectors were then pooled to form datasets corresponding to the UMi, UMa, and SMa scenarios.

\subsection{Measurement Hardware} 

The transmitter (TX) incorporates an OFDM-modulated Zadoff-Chu sequence with CP addition and time-domain repetitions. This modulated signal is up-converted to an RF frequency of  6.9, 8.3 and 14.5 GHz, prior to amplification transmission through a standard gain horn antenna. The transmitter delivers a maximum effective isotropic radiated power (EIRP) of +43 dBm over 400 MHz of instantaneous bandwidth.

At the receiver (RX), whether using the omnidirectional antenna or the phased array configuration, the RF signal is downconverted to a 3 GHz intermediate frequency (IF) and then to baseband. It is then demodulated using an OFDM receiver and correlated with the transmitted Zadoff-Chu (ZC) sequence to obtain the complex-valued channel impulse response (CIR), which can be recorded at speeds of up to 25 GB/s. The transmitter and receiver are synchronized using a GNSS-derived timing reference. Tables~\ref{tab:transmitter} and \ref{tab:receiver} provide the specifications for the receiver and transmitter, respectively.  Finally, the sounder was fully calibrated before and after each measurement campaign for accuracy and consistency of measured data. See~\cite{Nieman_2025_01} for additional details about this channel sounder.

\begin{table}[t!]
\vspace{0.03in}
	\caption{Transmitter Specifications}
	\begin{center}
		\begin{tabular}{| l | c |}
			\hline
			\textbf{Attributes} & \textbf{Values}\\
			\hline \cline{1-2}
			Frequency (GHz) & 6.9,8.3,14.5 \\
						\hline
			Bandwidth (MHz) & 400 \\
									\hline
			Modulation & OFDM \\
									\hline
			Subcarrier spacing (kHz) & 120 \\
									\hline
			ZC Sequence Length & 3343\\
									\hline
			Antenna, Gain (dBi) & Standard horn, 10  \\
									\hline
			Antenna 3-dB beamwidth & AZ: 55\textdegree, EL: 55\textdegree \\
									\hline
			Polarization & Vertical\\
									\hline
			Maximum EIRP (dBm) & +43\\
									\hline
			
			\hline
		\end{tabular}
		\label{tab:transmitter}
	\end{center}
\end{table}
\begin{table}[t!]
		\caption{Receiver Specifications}
		\begin{center}
			\begin{tabular}{| l | c |c |}
				\hline
				\textbf{Attributes} & \multicolumn{2}{|c|}{\textbf{Values}} \\
				\hline \cline{1-2}
				Antenna System& Omnidirectional & Phased Array \\
								\hline
				Frequency (GHz) & 6.9,8.3,14.5 & 8.3,14.5 \\
								\hline
				Min. MPC Delay Resolution (ns) & \multicolumn{2}{|c|}{2.5} \\
								\hline
				No. of antenna elements & (2) Omni\footnotemark[1] & 32 or 64\footnotemark[2]\\
								\hline
				No. of beams per array (8/15GHz) & N/A &  15/20 \\
								\hline
                Field of view &  \multicolumn{2}{|c|}{AZ: {360\textdegree}, EL:{$\pm$32.5\textdegree}} \\
                				\hline
				Receiver type & \multicolumn{2}{|c|}{Full correlator}\\
								\hline
                    Processing Gain (dB) & \multicolumn{2}{|c|}{41}\\
								\hline
				No. of receivers & 1 & 4 (1 per array) \\
								\hline
				360\textdegree\  CIR acquisition time & $<$ 40 $\mu$s& 0.5-0.9 ms\\
								\hline

				No. of averaging & \multicolumn{2}{|c|}{User selectable} \\
								\hline
				Max. Meas. Excess Delay& \multicolumn{2}{|c|}{8 $\mu$s} \\   		
						\hline
				Receiver noise figure& 1.5 dB & 8.3 dB \\
				\hline
                \multicolumn{3}{l}{\scriptsize$^1$Two Omni antennas may be used with any phase array frequency.}\\
                \multicolumn{3}{l}{\scriptsize$^2$32 elements for 8GHz array and 64 for 15 GHz array.}\\
			\end{tabular}
    			\label{tab:receiver}
		\end{center}
 
	\end{table}
\begin{figure}[!t]
    \centering
    \subfloat[Suburban-Macro Site]{
        \includegraphics[width=0.45\textwidth]{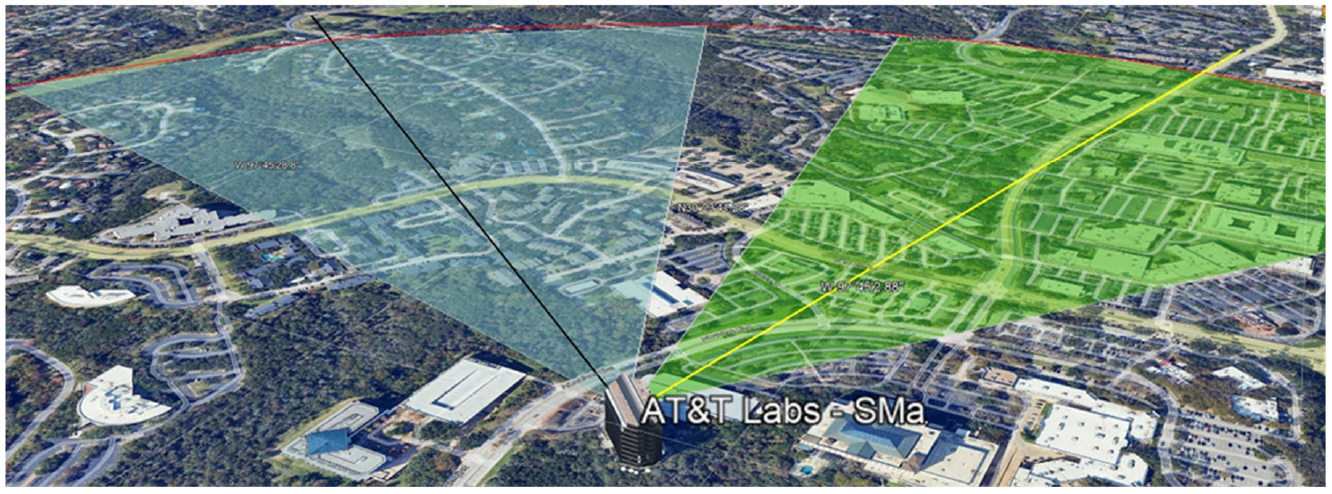} 
        \label{fig:SMa_Area}
    } \\
    \subfloat[Urban-Micro Site]{
        \includegraphics[width=0.45\textwidth]{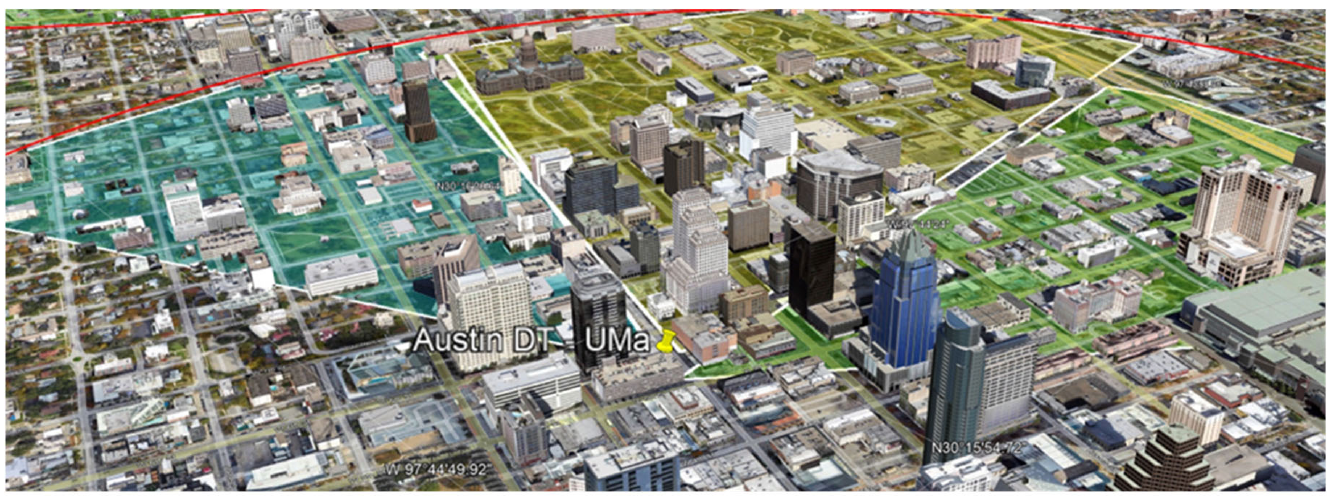} 
        \label{fig:UMa_Area}
    }
    \caption{SMa (a) and UMi (b) experiment coverage areas.}
    \label{fig:Depl_Scen}
    
\end{figure}
\subsection{Measurement Scenarios} 
\label{sec:experiment}
Measurement scenarios were modeled after those described in~\cite{3gpp.38.901}. UMi deployment scenarios represent areas with high building density, narrow street and small open spaces, significant clutter, high population density and varying building materials. UMa features varying building and clutter heights, wide streets and avenues, high population density, dense infrastructure and varying building materials. SMa is best characterized by moderate building density with building heights of one to four stories, significant amount of vegetation which can rise to about 25 m, lower population density, wide streets and larger lots. 

For UMi, the selected transmit location was a parking garage located between many high rise buildings in downtown Austin, Texas, USA. LOS conditions were possible along adjacent roadways with NLOS possible along subsequent grid of streets. For UMa, the transmitter was placed atop a parking garage located within the University of Texas at Austin campus. LOS conditions were very limited due to narrow roadways and abundant foliage. For SMa, two transmit locations were selected. The first was a rooftop deployment on a commercial building in the center of a suburban shopping mall. The second was a parking structure adjacent to a large residential neighborhood. With the exception of the UMi case, all deployment scenarios were selected such that the transmission site was above the clutter height. When possible, the transmitting antennas were collocated with existing 5G C-band radio installations, enabling future coverage comparison studies.  Table~\ref{tab:Scenarios} summarizes the features of the three deployment scenarios evaluated in this study. Figure~\ref{fig:Depl_Scen} shows an SMa and UMi sites with their respective coverage areas.

\begin{table}[t]
\centering

\caption{Deployment Scenarios}
\begin{tabular}{|c|cc|}
\hline
\textbf{Deployment Scenarios}               & \multicolumn{2}{c|}{\textbf{Specification}}              \\ \hline \cline{1-3}
\multirow{4}{*}{\textbf{Urban Micro (UMi)}} & \multicolumn{1}{c|}{ Vegetation height (m)}  & 2-20              \\ \cline{2-3} 
                                            & \multicolumn{1}{c|}{ Building height (m)} & 2-200          \\ \cline{2-3} 
                                            & \multicolumn{1}{c|}{Street length (m)}            & 550 \\ \cline{2-3} 
                                            & \multicolumn{1}{c|}{Street width (m)}            & 4-10              \\ \hline \cline{1-3}
\multirow{4}{*}{\textbf{Urban Macro (UMa)}}       & \multicolumn{1}{c|}{ Vegetation height (m)}  & 1-22              \\ \cline{2-3} 
                                            & \multicolumn{1}{c|}{ Building height (m)} & 2-90          \\ \cline{2-3} 
                                            & \multicolumn{1}{c|}{Street length (m)}            & 100-400 \\ \cline{2-3} 
                                            & \multicolumn{1}{c|}{Street width (m)}            & 4-10              \\ \hline \cline{1-3}
\multirow{4}{*}{\textbf{Suburban Macro (SMa)}}    & \multicolumn{1}{c|}{ Vegetation height (m)}  & 1-25              \\ \cline{2-3} 
                                            & \multicolumn{1}{c|}{ Building height (m)} & 2-35          \\ \cline{2-3} 
                                            & \multicolumn{1}{c|}{Street length (m)}            & 200-700 \\ \cline{2-3} 
                                            & \multicolumn{1}{c|}{Street width (m)}            & 4-10              \\ \hline \cline{1-3}                                
\end{tabular}
\label{tab:Scenarios}
\end{table}

 \subsection{Data Processing}
 \label{sec:data_processing}

Since the data capture rate at the RX was fixed, more data points were collected at locations where the RX van was moving slowly or stopped (e.g., at a traffic light). To ensure uniform data collection across all scenarios, the measured RX data were binned into 2-meter square bins. The measurement with the median path loss within each bin was then chosen as the representative measurement for evaluation of all parameters.

The omnidirectional path loss ($PL$) is modeled as:
\begin{equation} \label{PL-Eq}
  PL(d) = PL_0 + 10 PLE\; \log_{10} \left( \frac{d}{d_0} \right) + S \:[dB],
\end{equation}
where $d$ is the TR separation distance in m, $PL_{\text{0}}$ is the path loss intercept at $d_{\text{0}}$ = 100 m, \textit{PLE} is the path loss exponent of the channel, and $S$ is the shadow fading component. 

The RMS delay spread ($\tau_{rms}$) was then computed using data collected by the omnidirectional antenna at 6.9 GHz, as well as data collected by the phased arrays at 8.3 and 14.5 GHz. The mean delay ($\tau_{m}$) and $\tau_{rms}$ are given by:
\begin{equation} \label{mean_dly}
  \tau_{m} = \sqrt{\frac{\sum_{i=1}^{N}\tau_{i}P_i}{{\sum_{i=1}^{N}P_i}}}, \nonumber
\end{equation}
\begin{equation} \label{rms_dly}
   \tau_{rms} = \sqrt{\frac{\sum_{i=1}^{N}(\tau_{i}-\tau_{m}^{2})P_i}{\sum_{i=1}^{N}P_i}},\nonumber
\end{equation}
where $N$ is the number of channel taps, $\tau_i$ is the delay and $P_i$ is the power of the $i^{th}$ channel tap \cite{Rappaport_2002}.

We then examined coherence bandwidth ($B_c$) which is a measure of the frequency range over which the channel response is flat. This parameter helps determine the extent of frequency-selective fading and guides reference signal density. It may be calculated directly from the RMS delay spread as follows:

\begin{equation} \label{Cohere_BW}
 B_{c, \rho} \approx \frac{1}{K \tau_{rms}},\nonumber
\end{equation}
where $K$ is approximated as 5 or 50, if coherence bandwidth is selected such that the frequency correlation, $\rho$ is above 0.5 or 0.9, respectively \cite{Rappaport_2002}.

Angular spread ($AS$) is a measure of the spread of the power of the channel over angle and is calculated as: 
\begin{equation} \label{eq:as}
   AS = \sqrt{-2\ln\left(\left|\frac{\sum^{N}_{i=1}{e^{j\phi_{i}}}P_{i}}{\sum^{N}_{i=1}{P_{i}}}\right)\right|},\nonumber
\end{equation}
where $N$ is the number of channel taps, and $\phi_i$ is the angle channel tap $i$. $AS$ is defined at both the TX and the RX as the departure and arrival $AS$, respectively. In this work, using data collected by the phased arrays at 8.3, and 14.5 GHz, we computed the azimuth angle spread of arrival ($ASA$) and zenith angle spread of arrival ($ZSA$) as the spread of MPC power observed at the RX in azimuth and zenith dimensions, respectively.

\section{Experimental Results}
\label{sec:KeyFindings}

We now present analyses of large-scale fading, RMS delay spread/coherence bandwidth, and angular spread for the three deployment scenarios, along with their cross-correlation. The full set of large-scale channel parameters is listed in Table~\ref{tab:Ch_Params}. Due to space constraints, plots are shown only for the 15 GHz data (Figs.~\ref{fig:SF}–\ref{fig:ZSA}); results are representative of all measured frequencies.

\subsection{Large-Scale Fading}
\label{sec:fading}

\begin{figure}[t!]
    \centering
    \subfloat[LOS conditions]{
    \label{fig:UMi_PL_LOS}
        \includegraphics[width=0.97\columnwidth]{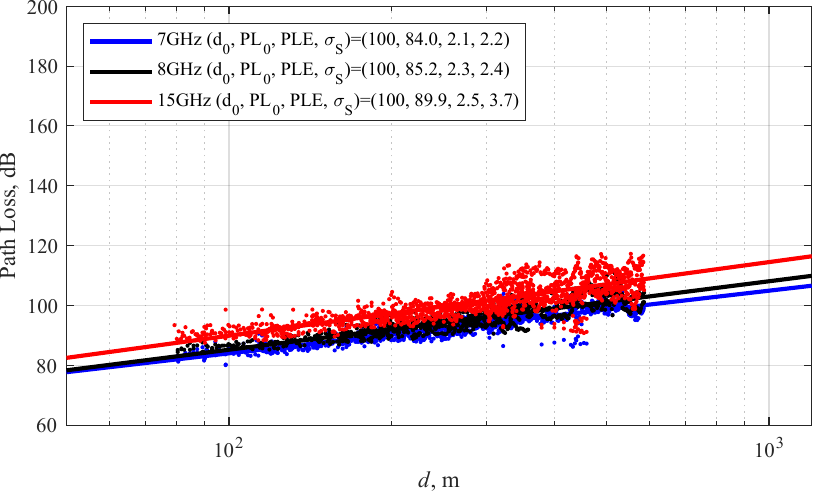} 
    } \\
    \subfloat[NLOS conditions]{
    \label{fig:UMi_PL_NLOS}
        \includegraphics[width=0.97\columnwidth]{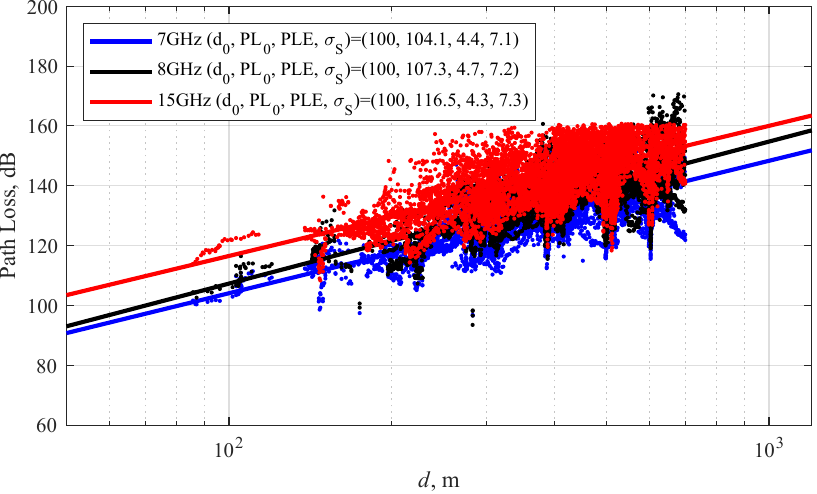} 
    }
    \caption{UMi}
    \label{fig:UMi_PL}
\end{figure}

Fig.~\ref{fig:UMi_PL} shows the path loss for the UMi scenario. Under LOS conditions (see Fig.~\ref{fig:UMi_PL_LOS}), we observe $PL_0$ of 84.0, 85.2, and 89.9 dB which are slightly lower than that of FSPL (88.9, 90.7, 95.6 dB), possibly due to waveguiding effect of street canyon. We observe $PLE$ of 2.1, 2.3, and 2.5 which are slightly higher than the 2.1 of~\cite{3gpp.38.901}. Shadow fading is 2.2 to 3.7 dB with higher shadowing observed in the higher frequency band. For NLOS UMi (see Fig.~\ref{fig:UMi_PL_NLOS}), we observe $PL_0$ of 101.7, 106.9, and 115.8 dB with $PLE$ of 4.3, 4.6, and 4.3 for 6.9, 8.3, and 14.5 GHz, respectively. Shadow fading $\sigma_S$ was measured as 7.1, 7.3, and 7.3 dB for the three frequencies.

\begin{figure}[t!]
    \centering

        \includegraphics[width=0.97\columnwidth]{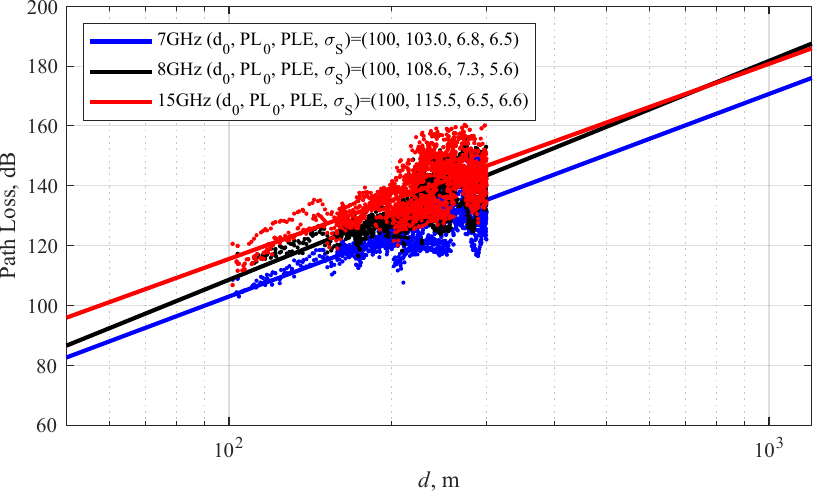} 
    \caption{UMa NLOS}
    \label{fig:UMa_PL}
\end{figure}

For UMa scenario, only NLOS data was collected. Fig.~\ref{fig:UMa_PL} shows path loss vs. $d$. Using $d_0$ of 100 m, we measure $PL_0$ of 103.0, 108.6, and 115.6 dB for 6.9, 8.3, 14.5 GHz, respectively. $PLE$ of 6.8, 7.3, and 6.5 and $\sigma_S$ of 6.5, 5.6, and 6.6 dB were observed. Path loss for this scenario was much higher than UMi. Key differences between UMa and UMi environment that could contribute to higher loss are: (\textit{i}) narrower vs. wider streets, (\textit{ii}) denser vs. lighter foliage, and (\textit{iii}) different building construction (e.g. thick concrete and masonry exteriors vs. glass and steel exteriors).  

\begin{figure}[t!]
    \centering
    \subfloat[LOS conditions]{
    \label{fig:SMa_PL_LOS}
        \includegraphics[width=0.97\columnwidth]{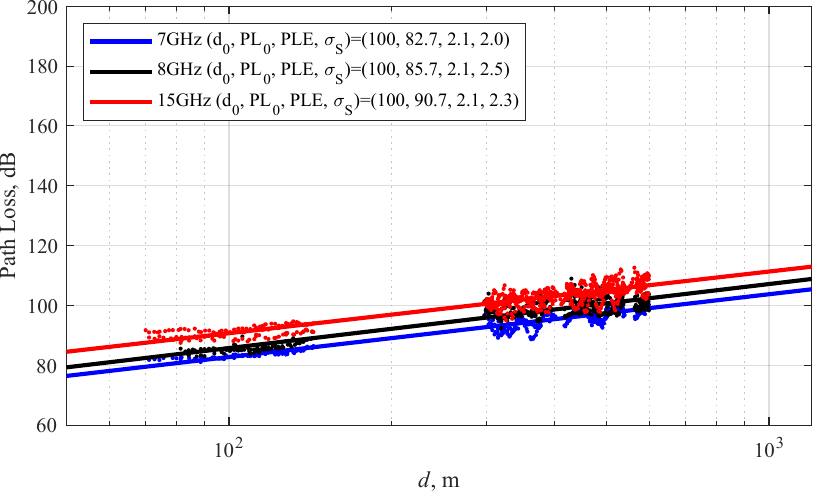} 
    } \\
    \subfloat[NLOS conditions]{
    \label{fig:SMa_PL_NLOS}
        \includegraphics[width=0.97\columnwidth]{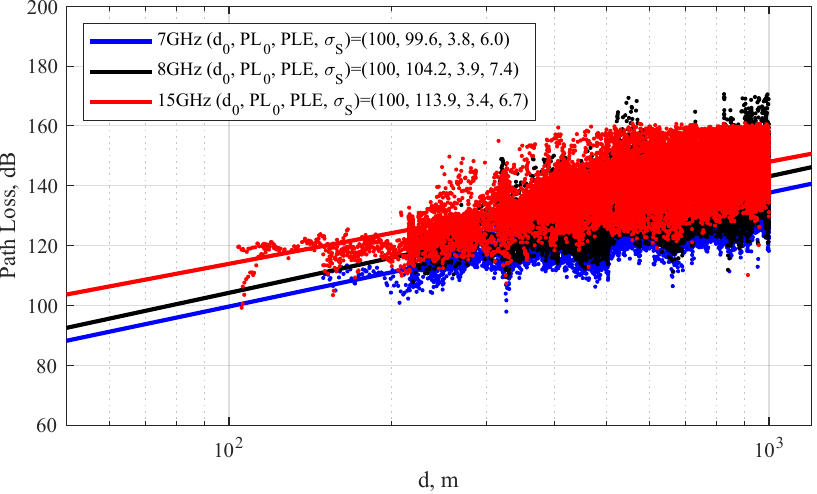} 
    }
    \caption{SMa}
    \label{fig:SMa_PL}
\end{figure}

Fig.~\ref{fig:SMa_PL} shows the path loss from SMa scenario. In LOS, we observe $PL_0$ of 43.8, 43.4, and 53.4 dB with PLE of 2.0, 2.1, 1.9, and a $\sigma_S$ of 2.3, 2.5, and 2.5 dB for 7, 8, 15 GHz, respectively. The portion of LOS data between 150 and 300 m was removed from the dataset due to foliage. In NLOS, we observe $PL_0$ of 23.5, 26.3, and 45.7 with PLE of 2.0, 2.1, 1.9, and a $\sigma_S$ of 2.3, 2.5, and 2.5 for 6.9, 8.3, 14.5 GHz, respectively.

\begin{figure}[t]
    \centering
    \subfloat[SF vs. $d$]{
  		\includegraphics[width=0.97\columnwidth]{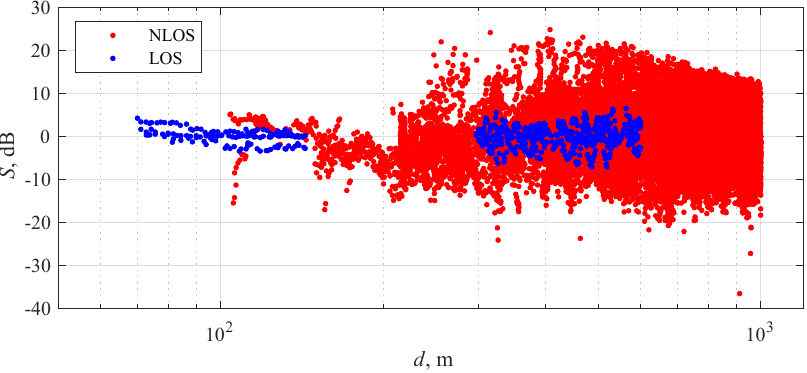}
		\label{fig:SF_vs_d}
        } \\  
    \subfloat[Probability plot of SF]{
  		\includegraphics[width=0.97\columnwidth]{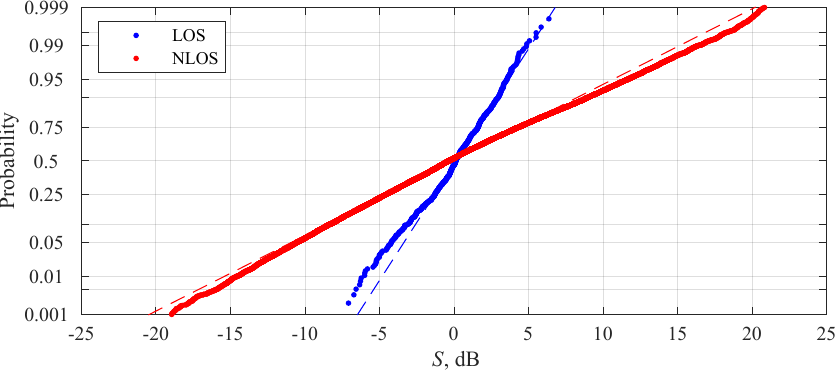}
		\label{fig:prob_plot_SF}
        }
    \caption{SF for SMa 14.5 GHz}
        \label{fig:SF}
\end{figure}

Fig.~\ref{fig:SF_vs_d} illustrates the shadow fading parameter $S$ as a function of $d$ at 15 GHz for SMa. $\sigma_S$ appears to increase positively with $d$. $S$ is typically modeled as a zero mean log-normal random variable. Fig.~\ref{fig:prob_plot_SF} confirms that $S$ has a log-normal distribution. $S$ closely follows a log-normal distribution across the other measured frequencies.

\subsection{RMS Delay Spread and Coherence Bandwidth}
\label{sec:delay}

Table~\ref{tab:Ch_Params} shows the mean and standard deviation of $\log_{10}(\tau_{rms}/1\text{s})$ for the three deployment scenarios in LOS and NLOS conditions. As expected, $\tau_{rms,\text{NLOS}}$ is larger than $\tau_{rms,\text{LOS}}$ by a factor of 4-10 in UMi and 3 in SMa. This difference is due to more dispersed multipath in NLOS conditions. When comparing UMi/UMa to SMa, we observe similar delay spread for LOS but greater delay spread for NLOS. Specifically, we observe a $\sim3\times$ greater delay spread for UMi/UMa vs. SMa in NLOS conditions. 

\begin{figure}[t]
    \centering
    \subfloat[DS vs. $d$]{
  		\includegraphics[width=0.97\columnwidth]{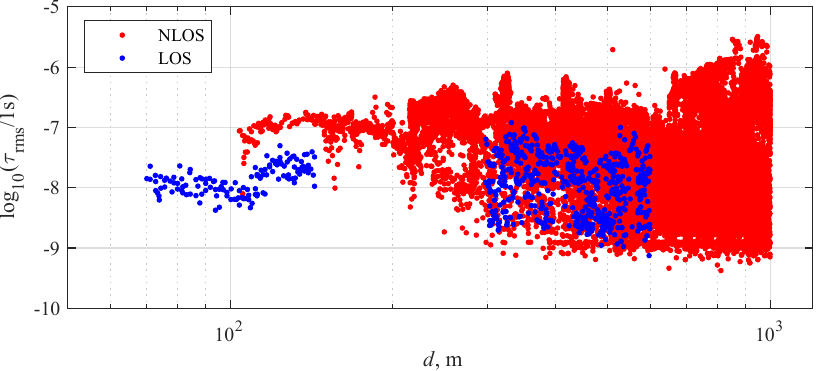}
		\label{fig:DS_vs_d}
        } \\  
    \subfloat[Probability plot of DS]{
  		\includegraphics[width=0.97\columnwidth]{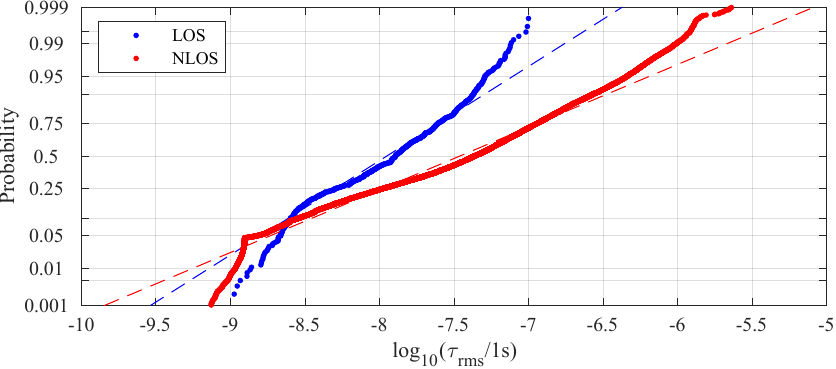}
		\label{fig:prob_plot_DS}
        }
    \caption{DS for SMa 14.5 GHz}
        \label{fig:DS}
\end{figure}

Fig.~\ref{fig:DS} shows typical results for delay spread vs. $d$ as well as the fit to log-normal distribution. Due to space constraints, we limit this detailed analysis to SMa 15 GHz measurements. These results are typical for other scenarios and frequencies. As seen in Fig.~\ref{fig:DS_vs_d}, delay spread variance appears to increase with increased $d$. Large delay spreads on the order of 1 $\mu\text{s}$) are observed beyond 700 m. Fig.~\ref{fig:prob_plot_DS} shows a good fit to log-normal in the 10-90\% region for LOS and 5-95\% region for NLOS. Measurement system limitation of finite bandwidth (400 MHz) and absolute delay (8 $\mu$s) contribute to the shape of the lower and upper tail probabilities, respectively. 

Our measurements indicate small 90\% coherence bandwidths of 0.1-0.2 MHz, particularly in UMi/UMa NLOS scenarios. This implies that frequency-domain reference signal spaced at least this close to allow for coherent detection over these channels~\cite{Chung_2022}. 

\setlength{\tabcolsep}{2.1pt}
\begin{table*}[!t]
\caption{Outdoor Channel Parameters}
\label{tab:Ch_Params}
\centering

\begin{tabular}{|cc|cccccc|cccccc|cccccc|}
\hline
\multicolumn{2}{|c|}{\multirow{3}{*}{Parameters}} & \multicolumn{6}{c|}{7 GHz} & \multicolumn{6}{c|}{8 GHz} & \multicolumn{6}{c|}{15 GHz} \\ \cline{3-20}  \cline{3-20}
\multicolumn{2}{|c|}{} & \multicolumn{2}{c|}{UMi} & \multicolumn{2}{c|}{UMa} & \multicolumn{2}{c|}{SMa} & \multicolumn{2}{c|}{UMi} & \multicolumn{2}{c|}{UMa} & \multicolumn{2}{c|}{SMa} & \multicolumn{2}{c|}{UMi} & \multicolumn{2}{c|}{UMa} & \multicolumn{2}{c|}{SMa} \\ \cline{3-20} \cline{3-20}

\multicolumn{2}{|c|}{} & \multicolumn{1}{c|}{LOS} & \multicolumn{1}{c|}{NLOS} & \multicolumn{1}{c|}{LOS} & \multicolumn{1}{c|}{NLOS} & \multicolumn{1}{c|}{LOS} & NLOS & \multicolumn{1}{c|}{LOS} & \multicolumn{1}{c|}{NLOS} & \multicolumn{1}{c|}{LOS} & \multicolumn{1}{c|}{NLOS} & \multicolumn{1}{c|}{LOS} & NLOS & \multicolumn{1}{c|}{LOS} & \multicolumn{1}{c|}{NLOS} & \multicolumn{1}{c|}{LOS} & \multicolumn{1}{c|}{NLOS} & \multicolumn{1}{c|}{LOS} & NLOS \\ \hline \cline{1-20}

\multicolumn{1}{|c|}{\multirow{4}{*}{Path loss, dB}} & $PL_0$, dB & \multicolumn{1}{c|}{84.6} & \multicolumn{1}{c|}{104.4} & \multicolumn{1}{c|}{--} & \multicolumn{1}{c|}{103.0} & \multicolumn{1}{c|}{45.9} & 99.6 & \multicolumn{1}{c|}{86.2} & \multicolumn{1}{c|}{108.0} & \multicolumn{1}{c|}{--} & \multicolumn{1}{c|}{108.6} & \multicolumn{1}{c|}{40.1} & 104.2 & \multicolumn{1}{c|}{90.2} & \multicolumn{1}{c|}{116.6} & \multicolumn{1}{c|}{--} & \multicolumn{1}{c|}{115.6} & \multicolumn{1}{c|}{50.8} & 45.7 \\ \cline{2-20}  

\multicolumn{1}{|c|}{} & $d_0$, m & \multicolumn{2}{c|}{100} & \multicolumn{2}{c|}{100} & \multicolumn{2}{c|}{100} & \multicolumn{2}{c|}{100} & \multicolumn{2}{c|}{100} & \multicolumn{2}{c|}{100} & \multicolumn{2}{c|}{100} & \multicolumn{2}{c|}{100} & \multicolumn{2}{c|}{100} \\ \cline{2-20} 

\multicolumn{1}{|c|}{} & $PLE$ & \multicolumn{1}{c|}{2.1} & \multicolumn{1}{c|}{4.3} & \multicolumn{1}{c|}{--} & \multicolumn{1}{c|}{6.8} & \multicolumn{1}{c|}{1.9} & 3.8 & \multicolumn{1}{c|}{2.2} & \multicolumn{1}{c|}{4.6} & \multicolumn{1}{c|}{--} & \multicolumn{1}{c|}{7.3} & \multicolumn{1}{c|}{2.3} & 3.9 & \multicolumn{1}{c|}{2.5} & \multicolumn{1}{c|}{4.3} & \multicolumn{1}{c|}{--} & \multicolumn{1}{c|}{6.5} & \multicolumn{1}{c|}{2.0} & 3.4 \\ \cline{2-20} 

\multicolumn{1}{|c|}{} & $\sigma_S$, dB & \multicolumn{1}{c|}{2.4} & \multicolumn{1}{c|}{7.1} & \multicolumn{1}{c|}{--} & \multicolumn{1}{c|}{6.5} & \multicolumn{1}{c|}{2.5} & 6.0 & \multicolumn{1}{c|}{2.8} & \multicolumn{1}{c|}{7.3} & \multicolumn{1}{c|}{--} & \multicolumn{1}{c|}{5.6} & \multicolumn{1}{c|}{2.9} & 7.4 & \multicolumn{1}{c|}{3.3} & \multicolumn{1}{c|}{7.3} & \multicolumn{1}{c|}{--} & \multicolumn{1}{c|}{6.6} & \multicolumn{1}{c|}{2.9} & 6.7 \\ \hline \cline{1-20}

\multicolumn{1}{|c|}{\multirow{2}{*}{$\log_{10}$(DS/1s)}} & $\mu$ & \multicolumn{1}{c|}{-7.61} & \multicolumn{1}{c|}{-6.77} & \multicolumn{1}{c|}{--} & \multicolumn{1}{c|}{-6.77} & \multicolumn{1}{c|}{-7.75} &  -7.20& \multicolumn{1}{c|}{-7.65} & \multicolumn{1}{c|}{-6.63} & \multicolumn{1}{c|}{--} & \multicolumn{1}{c|}{-6.81} & \multicolumn{1}{c|}{-7.67} & -7.18 & \multicolumn{1}{c|}{-7.65} & \multicolumn{1}{c|}{-7.04} & \multicolumn{1}{c|}{--} & \multicolumn{1}{c|}{-7.03} & \multicolumn{1}{c|}{-7.94} &  -7.46\\ \cline{2-20} 

\multicolumn{1}{|c|}{} & $\sigma$ & \multicolumn{1}{c|}{0.44} & \multicolumn{1}{c|}{0.70} & \multicolumn{1}{c|}{--} & \multicolumn{1}{c|}{0.38} & \multicolumn{1}{c|}{0.36} & 0.59 & \multicolumn{1}{c|}{0.44} & \multicolumn{1}{c|}{0.63} & \multicolumn{1}{c|}{--} & \multicolumn{1}{c|}{0.41} & \multicolumn{1}{c|}{0.36} & 0.58 & \multicolumn{1}{c|}{0.44} & \multicolumn{1}{c|}{0.79} & \multicolumn{1}{c|}{--} & \multicolumn{1}{c|}{0.50} & \multicolumn{1}{c|}{0.45} &  0.73\\ \hline \cline{1-20}

\multicolumn{1}{|c|}{\multirow{2}{*}{$C_{b,\rho}$, MHz}} & $\rho =0.5$ & \multicolumn{1}{c|}{8.1} & \multicolumn{1}{c|}{1.2} & \multicolumn{1}{c|}{--} & \multicolumn{1}{c|}{1.2} & \multicolumn{1}{c|}{11.2} &  3.2& \multicolumn{1}{c|}{8.9} & \multicolumn{1}{c|}{0.9} & \multicolumn{1}{c|}{--} & \multicolumn{1}{c|}{1.3} & \multicolumn{1}{c|}{9.4} & 3.0 & \multicolumn{1}{c|}{8.9} & \multicolumn{1}{c|}{2.2} & \multicolumn{1}{c|}{--} & \multicolumn{1}{c|}{2.1} & \multicolumn{1}{c|}{17.4} &  5.8\\ \cline{2-20} 

\multicolumn{1}{|c|}{} & $\rho =0.9$ & \multicolumn{1}{c|}{0.8} & \multicolumn{1}{c|}{0.1} & \multicolumn{1}{c|}{--} & \multicolumn{1}{c|}{0.1} & \multicolumn{1}{c|}{1.1} & 0.3 & \multicolumn{1}{c|}{0.9} & \multicolumn{1}{c|}{0.1} & \multicolumn{1}{c|}{--} & \multicolumn{1}{c|}{0.1} & \multicolumn{1}{c|}{0.9} & 0.3 & \multicolumn{1}{c|}{0.9} & \multicolumn{1}{c|}{0.2} & \multicolumn{1}{c|}{--} & \multicolumn{1}{c|}{0.2} & \multicolumn{1}{c|}{1.7} &  0.6\\ \hline \cline{1-20}

\multicolumn{1}{|c|}{\multirow{2}{*}{$\log_{10}$(ASA/1\textdegree)}} & $\mu$ & \multicolumn{1}{c|}{--} & \multicolumn{1}{c|}{--} & \multicolumn{1}{c|}{--} & \multicolumn{1}{c|}{--} & \multicolumn{1}{c|}{--} & -- & \multicolumn{1}{c|}{1.62} & \multicolumn{1}{c|}{1.57} & \multicolumn{1}{c|}{--} & \multicolumn{1}{c|}{1.71} & \multicolumn{1}{c|}{1.75} & 1.69 & \multicolumn{1}{c|}{1.29} & \multicolumn{1}{c|}{1.34} & \multicolumn{1}{c|}{--} & \multicolumn{1}{c|}{1.44} & \multicolumn{1}{c|}{1.07} & 1.51 \\ \cline{2-20} 

\multicolumn{1}{|c|}{} & $\sigma$ & \multicolumn{1}{c|}{--} & \multicolumn{1}{c|}{--} & \multicolumn{1}{c|}{--} & \multicolumn{1}{c|}{--} & \multicolumn{1}{c|}{--} & -- & \multicolumn{1}{c|}{0.05} & \multicolumn{1}{c|}{0.24} & \multicolumn{1}{c|}{--} & \multicolumn{1}{c|}{0.18} & \multicolumn{1}{c|}{0.45} & 0.51 & \multicolumn{1}{c|}{0.15} & \multicolumn{1}{c|}{0.37} & \multicolumn{1}{c|}{--} & \multicolumn{1}{c|}{0.30} & \multicolumn{1}{c|}{0.74} &  0.34\\ \hline \cline{1-20}

\multicolumn{1}{|c|}{\multirow{2}{*}{$\log_{10}$(ZSA/1\textdegree)}} & $\mu$ & \multicolumn{1}{c|}{--} & \multicolumn{1}{c|}{--} & \multicolumn{1}{c|}{--} & \multicolumn{1}{c|}{--} & \multicolumn{1}{c|}{--} & -- & \multicolumn{1}{c|}{1.30} & \multicolumn{1}{c|}{1.26} & \multicolumn{1}{c|}{--} & \multicolumn{1}{c|}{1.32} & \multicolumn{1}{c|}{1.31} & 1.32 & \multicolumn{1}{c|}{0.91} & \multicolumn{1}{c|}{0.95} & \multicolumn{1}{c|}{--} & \multicolumn{1}{c|}{0.96} & \multicolumn{1}{c|}{1.04} & 1.03 \\ \cline{2-20} 

\multicolumn{1}{|c|}{} & $\sigma$ & \multicolumn{1}{c|}{--} & \multicolumn{1}{c|}{--} & \multicolumn{1}{c|}{--} & \multicolumn{1}{c|}{--} & \multicolumn{1}{c|}{--} & -- & \multicolumn{1}{c|}{0.02} & \multicolumn{1}{c|}{0.09} & \multicolumn{1}{c|}{--} & \multicolumn{1}{c|}{0.06} & \multicolumn{1}{c|}{0.02} & 0.09 & \multicolumn{1}{c|}{0.07} & \multicolumn{1}{c|}{0.18} & \multicolumn{1}{c|}{--} & \multicolumn{1}{c|}{0.22} & \multicolumn{1}{c|}{0.08} & 0.19 \\ \hline \cline{1-20}

\multicolumn{1}{|c|}{\multirow{6}{*}{\makecell{Correlation \\coefficients}}} & ASA vs DS & \multicolumn{1}{c|}{--} & \multicolumn{1}{c|}{--} & \multicolumn{1}{c|}{--} & \multicolumn{1}{c|}{--} & \multicolumn{1}{c|}{--} & -- & \multicolumn{1}{c|}{0.19} & \multicolumn{1}{c|}{0.39} & \multicolumn{1}{c|}{--} & \multicolumn{1}{c|}{-0.38} & \multicolumn{1}{c|}{0.61} & 0.50 & \multicolumn{1}{c|}{0.79} & \multicolumn{1}{c|}{0.46} & \multicolumn{1}{c|}{--} & \multicolumn{1}{c|}{0.49} & \multicolumn{1}{c|}{0.80} & 0.77 \\ \cline{2-20} 

\multicolumn{1}{|c|}{} & ASA vs SF & \multicolumn{1}{c|}{--} & \multicolumn{1}{c|}{--} & \multicolumn{1}{c|}{--} & \multicolumn{1}{c|}{--} & \multicolumn{1}{c|}{--} & -- & \multicolumn{1}{c|}{-0.12} & \multicolumn{1}{c|}{0.32} & \multicolumn{1}{c|}{--} & \multicolumn{1}{c|}{0.27} & \multicolumn{1}{c|}{-0.30} & 0.25 & \multicolumn{1}{c|}{0.19} & \multicolumn{1}{c|}{0.23} & \multicolumn{1}{c|}{--} & \multicolumn{1}{c|}{0.13} & \multicolumn{1}{c|}{-0.66} & 0.21 \\ \cline{2-20} 

\multicolumn{1}{|c|}{} & DS vs SF & \multicolumn{1}{c|}{-0.57} & \multicolumn{1}{c|}{0.15} & \multicolumn{1}{c|}{--} & \multicolumn{1}{c|}{0.10} & \multicolumn{1}{c|}{-0.62} & 0.29 & \multicolumn{1}{c|}{0.08} & \multicolumn{1}{c|}{0.04} & \multicolumn{1}{c|}{--} & \multicolumn{1}{c|}{-0.18} & \multicolumn{1}{c|}{-0.45} & 0.13 & \multicolumn{1}{c|}{0.06} & \multicolumn{1}{c|}{0.17} & \multicolumn{1}{c|}{--} & \multicolumn{1}{c|}{0.06} & \multicolumn{1}{c|}{-0.27} & 0.17 \\ \cline{2-20} 

\multicolumn{1}{|c|}{} & ZSA vs SF & \multicolumn{1}{c|}{--} & \multicolumn{1}{c|}{--} & \multicolumn{1}{c|}{--} & \multicolumn{1}{c|}{--} & \multicolumn{1}{c|}{--} & -- & \multicolumn{1}{c|}{0.09} & \multicolumn{1}{c|}{0.49} & \multicolumn{1}{c|}{--} & \multicolumn{1}{c|}{0.67} & \multicolumn{1}{c|}{-0.29} & 0.47 & \multicolumn{1}{c|}{-0.38} & \multicolumn{1}{c|}{-0.14} & \multicolumn{1}{c|}{--} & \multicolumn{1}{c|}{0.48} & \multicolumn{1}{c|}{0.08} & -0.03 \\ \cline{2-20} 

\multicolumn{1}{|c|}{} & ZSA vs DS & \multicolumn{1}{c|}{--} & \multicolumn{1}{c|}{--} & \multicolumn{1}{c|}{--} & \multicolumn{1}{c|}{--} & \multicolumn{1}{c|}{--} & -- & \multicolumn{1}{c|}{-0.15} & \multicolumn{1}{c|}{0.04} & \multicolumn{1}{c|}{--} & \multicolumn{1}{c|}{-0.66} & \multicolumn{1}{c|}{0.12} & -0.06 & \multicolumn{1}{c|}{-0.22} & \multicolumn{1}{c|}{0.20} & \multicolumn{1}{c|}{--} & \multicolumn{1}{c|}{-0.10} & \multicolumn{1}{c|}{0.15} & -0.09 \\ \cline{2-20} 

\multicolumn{1}{|c|}{} & ZSA vs ASA & \multicolumn{1}{c|}{} & \multicolumn{1}{c|}{--} & \multicolumn{1}{c|}{--} & \multicolumn{1}{c|}{--} & \multicolumn{1}{c|}{--} & -- & \multicolumn{1}{c|}{0.40} & \multicolumn{1}{c|}{0.49} & \multicolumn{1}{c|}{--} & \multicolumn{1}{c|}{0.55} & \multicolumn{1}{c|}{0.16} & 0.36 & \multicolumn{1}{c|}{-0.09} & \multicolumn{1}{c|}{0.22} & \multicolumn{1}{c|}{--} & \multicolumn{1}{c|}{0.29} & \multicolumn{1}{c|}{0.09} & 0.06 \\ \hline \cline{1-20}

\multicolumn{2}{|c|}{Measured Points (thousands)} & \multicolumn{1}{c|}{1.0} & \multicolumn{1}{c|}{4.6} & \multicolumn{1}{c|}{--} & \multicolumn{1}{c|}{6.4} & \multicolumn{1}{c|}{0.7} & 21.2 & \multicolumn{1}{c|}{1.0} & \multicolumn{1}{c|}{5.7} & \multicolumn{1}{c|}{--} & \multicolumn{1}{c|}{6.8} & \multicolumn{1}{c|}{0.8} & 22.5 & \multicolumn{1}{c|}{1.5} & \multicolumn{1}{c|}{7.6} & \multicolumn{1}{c|}{--} & \multicolumn{1}{c|}{6.7} & \multicolumn{1}{c|}{1.2} & 24.7 \\ \hline 
\end{tabular}
\end{table*}

\subsection{Azimuth and Zenith Angle Spread of Arrival}
\label{sec:angle}
Table~\ref{tab:Ch_Params} shows the of mean and standard deviation of $\log_{10}(\text{ASA/1\textdegree})$ and $\log_{10}(\text{ZSA/1\textdegree})$ for the three deployment scenarios in LOS and NLOS conditions. Angular measurements were unavailable for 6.9 GHz since we did not have a phased array antenna for that frequency. For 8.3 GHz, we observe LOS $\log_{10}(\text{ASA/1\textdegree})$ of 1.62 and 1.75 for UMi and SMa. For 14.5 Ghz, we smaller angular spread of 1.29 and 1.07 for LOS UMi and SMa, respectively. For SMa 14.5 GHz, we observe ASA in NLOS conditions that is nearly 3$\times$ that of LOS. For UMi, LOS and NLOS ASA are much more comparable. SMa LOS has the narrowest angular spread of the deployment scenarios.

\begin{figure}[t]
    \centering
    \subfloat[ASA vs. $d$]{
  		\includegraphics[width=0.97\columnwidth]{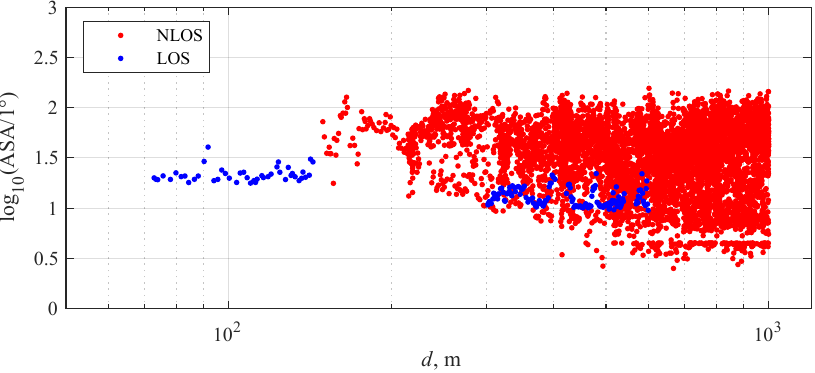}
		\label{fig:ASA_vs_d}
        } \\  
    \subfloat[Probability plot of ASA]{
  		\includegraphics[width=0.97\columnwidth]{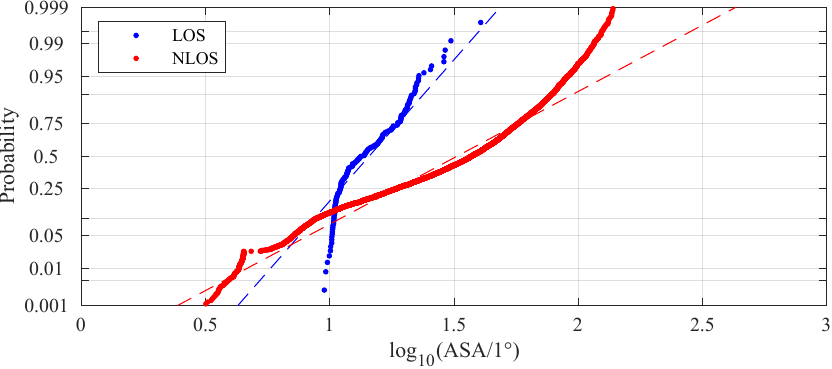}
		\label{fig:prob_plot_ASA}
        }
    \caption{ASA for SMa 14.5 GHz}
        \label{fig:ASA}
\end{figure}

Fig.~\ref{fig:ASA} shows ASA as a function of $d$ as well as its distribution vs. log-normal cumulative density function (CDF). As seen in Fig.~\ref{fig:ASA_vs_d}, ASA has a slight negative correlation with $d$ for LOS. This dependency leads to the shortened tail probability observed in Fig.~\ref{fig:prob_plot_ASA}. This effect appears to be due to the limited angular diversity of this particular channel: a long, straight road with adjacent 1-2 story buildings. A distance-dependent mean as was done in~\cite{Kanhere_2025_02} may provide a better fit to LOS data in this channel. For NLOS, ASA closely fits log-normal distribution below 90\% probability. The upper tail probability is affected by ambiguity of arrival angles beyond $\pm$180\textdegree due to circular wrapping of observed incident angle. In log domain this occurs at $\log_{10}\text{180\textdegree} = 2.26$. The circular wrapping is more pronounced on the upper tail of the distribution due to observed sample mean of 1.51.  

\begin{figure}[t]
    \centering
    \subfloat[ZSA vs. $d$]{
  		\includegraphics[width=0.97\columnwidth]{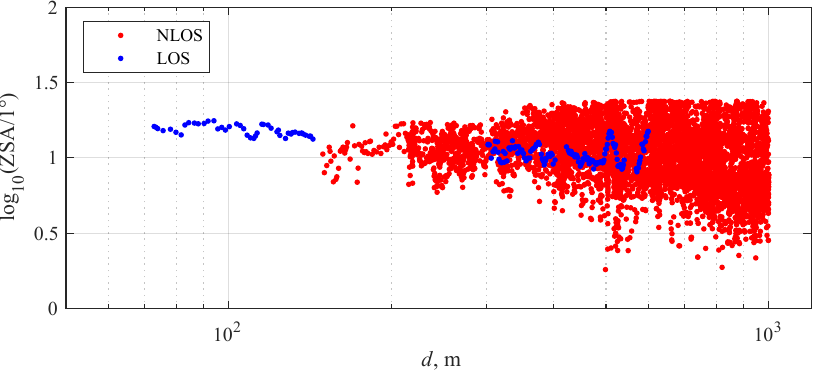}
		\label{fig:ZSA_vs_d}
        } \\  
    \subfloat[Probability plot of ZSA]{
  		\includegraphics[width=0.97\columnwidth]{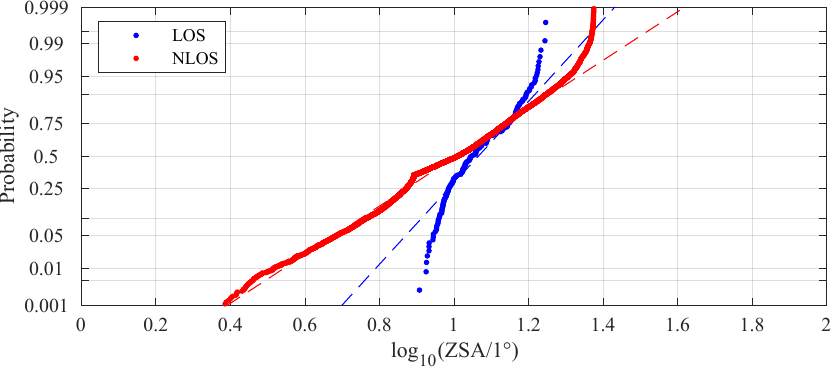}
		\label{fig:prob_plot_ZSA}
        }
    \caption{ZSA for SMa 14.5 GHz}
        \label{fig:ZSA}
\end{figure}

For ZSA we observe a similar dependence on $d$ for LOS data (see Fig.~\ref{fig:ZSA_vs_d}). This dependence leads to an ill-fit to log-normal distribution for LOS ZSA as shown in Fig.~\ref{fig:prob_plot_ZSA}. For NLOS, ZSA closely fits log-normal CDF below 95\%. Above 95\%, the data is limited by the maximum measurable $\log_{10}$ZSA of 1.37 due to the most extreme elevation pointing directions of $\pm23$\textdegree. This truncates the measurements and slightly skews the distribution from log-normal in the upper tail (large ZSA values). Since this is only a small portion of measured data, this does not significantly affect the reported mean or standard deviation..

\subsection{Inter-Parameter Cross-Correlations}
Table~\ref{tab:Ch_Params} lists cross-correlations of parameters with each other. For SMa scenario at 14.5 GHz, ASA vs. DS is correlated with mean cross-correlation of 0.80 and 0.77 for LOS and LOS, respectively. ASA vs. SF have a mean cross-correlation of -0.66 for LOS and 0.21 for NLOS. For DS vs. SF, we find a cross-correlation of -0.27 for LOS and  0.17. Looking at DS vs. SF cross-correlation for all frequencies measured, we observe a weakening trend with increasing frequency. Cross-correlations of ZSA and other parameters do not appear significant in the data collected. 

\section{Conclusions and Future Work}
\label{sec:conclusions}
In this paper we presented a comprehensive measurement and analysis of large-scale wireless channels in three outdoor deployment scenarios across multiple FR3 frequency bands. For each frequency band, key findings of the large-scale channel parameters were extracted and analyzed. The parameters included path loss, shadowing, RMS delay spread, and large-scale angular properties. We also investigated and reported on the cross-correlation between these parameters to identify relationships and dependencies. The insights gained from this analysis have been applied to refinement of outdoor wireless channel models within 3GPP and industry, with the ultimate goal of enhancing network design and performance in cellular networks. This work contributes to the optimization of next-generation wireless systems, especially in the FR3 spectrum.

\bibliographystyle{IEEEtran}
\bibliography{references}

\end{document}